\documentclass[12pt]{iopart}
\usepackage{latexsym}
\usepackage{fullpage}
\usepackage{graphics}
\usepackage{graphicx}

\begin{document}
\title{A torsion balance for probing a non-standard force in the sub-micrometre range}

\author{
M Masuda$^{1}$,
M Sasaki$^{1}$ and
A Araya$^{2}$
}

\address{
${}^{1}$Institute for Cosmic Ray Research, 
University of Tokyo, 5-1-5 Kashiwa-no-ha, Kashiwa-shi, Chiba 277-8582, Japan}
\address{
${}^{2}$Earthquake Research Institute, 
University of Tokyo, 1-1-1 Yayoi, Bunkyo-ku, Tokyo 113-0032, Japan
}

\ead{masuda@icrr.u-tokyo.ac.jp}

\begin{abstract}
We report the performance of an instrument that employs a torsion balance 
for probing a non-standard force in the sub-micrometre range. 
High sensitivity is achieved by using a torsion balance 
that has a long torsional period, strong magnetic damping of all vibrational motions and a feedback system that employs an optical lever. 
In torsion balance experiments, 
the distance fluctuations during measurements and 
the accuracy to which the absolute distance is determined are crucial 
for determining the sensitivity of the balance to a macroscopic force in the sub-micrometre range. 
We have estimated the root mean square amplitude of the distance fluctuation to be 18 nm 
by considering the effects due to seismic motions, tilt motions, residual angular fluctuations and 
thermal fluctuations.
We have also estimated the error of the absolute distance to be 13 nm 
and the statistical error of the force to be $3.4\times 10^{-12}$ N 
by measuring the electrostatic forces. 
As a result of this systematic study, we have evaluated the sensitivity of the balance to both 
a non-standard force and to the Casimir force. 
\end{abstract}


\section{Introduction}
\label{introduction} 
Quantum electrodynamics predicts that 
creation and annihilation of particle-antiparticle pairs constantly occur in vacuum 
and this microscopic phenomenon results in a macroscopic force between conductors, 
called the Casimir force \cite{Casimir}. 
We have focused on the fact that the existence of a macroscopic force that lies outside the standard model or Newtonian gravitational law should produce deviations from the Casimir force. 
Some exotic models which assume the existence of extra spatial dimensions or new particles 
predict non-standard forces in the range which the Casimir force is dominant \cite{Rubakov1}--\cite{Dimopoulos}. 
In order to probe such unknown forces, we have developed an instrument to search for deviations from the exact Casimir force.

The Casimir force, as predicted by standard quantum electrodynamics, 
between a flat conductor and a spherical conductor is expressed as 
\begin{eqnarray}
F_{cas}(d)=\frac{{\pi}^3{\hbar}c}{360}\frac{R}{d^3} \quad  \mathrm{for} \quad d\ll R,
\label{Radius}
\end{eqnarray}
where $d$ is the distance between the conductors, $\hbar$ is the Dirac constant,
$c$ is the speed of light and $R$ is the radius of the spherical lens. 

When calculating the Casimir force between real metals, 
corrections for the properties of the metals 
(such as finite conductivity, nonzero temperature and roughness of their surfaces) must be taken into account.
Accurate computations can be performed for the finite conductivity and the roughness corrections (see the review paper \cite{Bordag0}),  
however, four different models for the nonzero temperature 
correction have been proposed \cite{Bostrom1}--\cite{Bezerra}. 
Of these four models, two of them have been discounted on the basis of experimental measurements \cite{Decca0} while 
the discrepancy between the nonzero temperature correction calculated
using the remaining two models is much smaller than experimental accuracy. 
Thus, the existence of a non-standard force can be demonstrated by 
searching for deviations between theoretical and experimental Casimir forces.

In the last decade, various experimental demonstrations of the Casimir force have been reported. 
Small devices \cite{Mohideen1}--\cite{Bressi2} such as 
atomic force microscopes or micro-electro-mechanical systems 
have frequently been used to measure the Casimir force 
ever since Lamoreaux measured the Casimir force using a torsion balance \cite{Lamoreaux1}.
Torsion balances have several advantages over other devices
for probing non-standard forces. 
These include a high sensitivity to macroscopic forces, 
since a torsional period of the order of 1000 s is attainable, and 
the capability of using massive materials in torsion balance experiments.
This latter advantage is significant since many models predict that the non-standard force is  approximately proportional to the mass densities of the conductors \cite{Bordag2}, 
in addition, the sensitivity to the Casimir force is proportional to the radius $R$ of the spherical lens.
On the other hand, special care must be taken to ensure that the distance between the conductors remains constant when using torsion balances. 
This is not only because torsion balances are sensitive to 
disturbances such as seismic motions and ground tilts, 
but also because distance fluctuations produce large systematic errors
as a result of the strong dependence of the Casimir force on distance. 
Various noises, including thermal noise and detection noise, 
have been reported in torsion balance experiments \cite{Boynton}--\cite{Lamoreaux0}, 
however, there has never been a quantitative estimation of the extent to which seismic motions 
and ground tilts affect measurement of a macroscopic force in the sub-micrometre range. 

In this paper, we report the performance of an instrument that uses a torsion balance 
to search for non-standard forces in the sub-micrometre range. 
A feedback system with an optical lever is employed to keep the torsion angle fixed 
enabling us to measure the macroscopic force in the sub-micrometre range stably. 
We estimated the statistical errors in measuring the dependence of the force on the distance 
by measuring electric forces. 
We then estimated the distance fluctuations due to 
seismic motions, tilt motions, residual angular fluctuations and thermal fluctuations. 
In addition, the accuracy in determining the absolute distance was estimated by measuring electric forces. 
The effects of distance fluctuations and the error in determining the absolute distance 
are treated as systematic errors. 
Finally, we estimate the sensitivity to the Casimir force and the non-standard force from this systematic study.

\section{Instrument}
\label{sec:instrument}
\begin{figure}[h]
    \begin{center} 
    \includegraphics[width=12cm]
{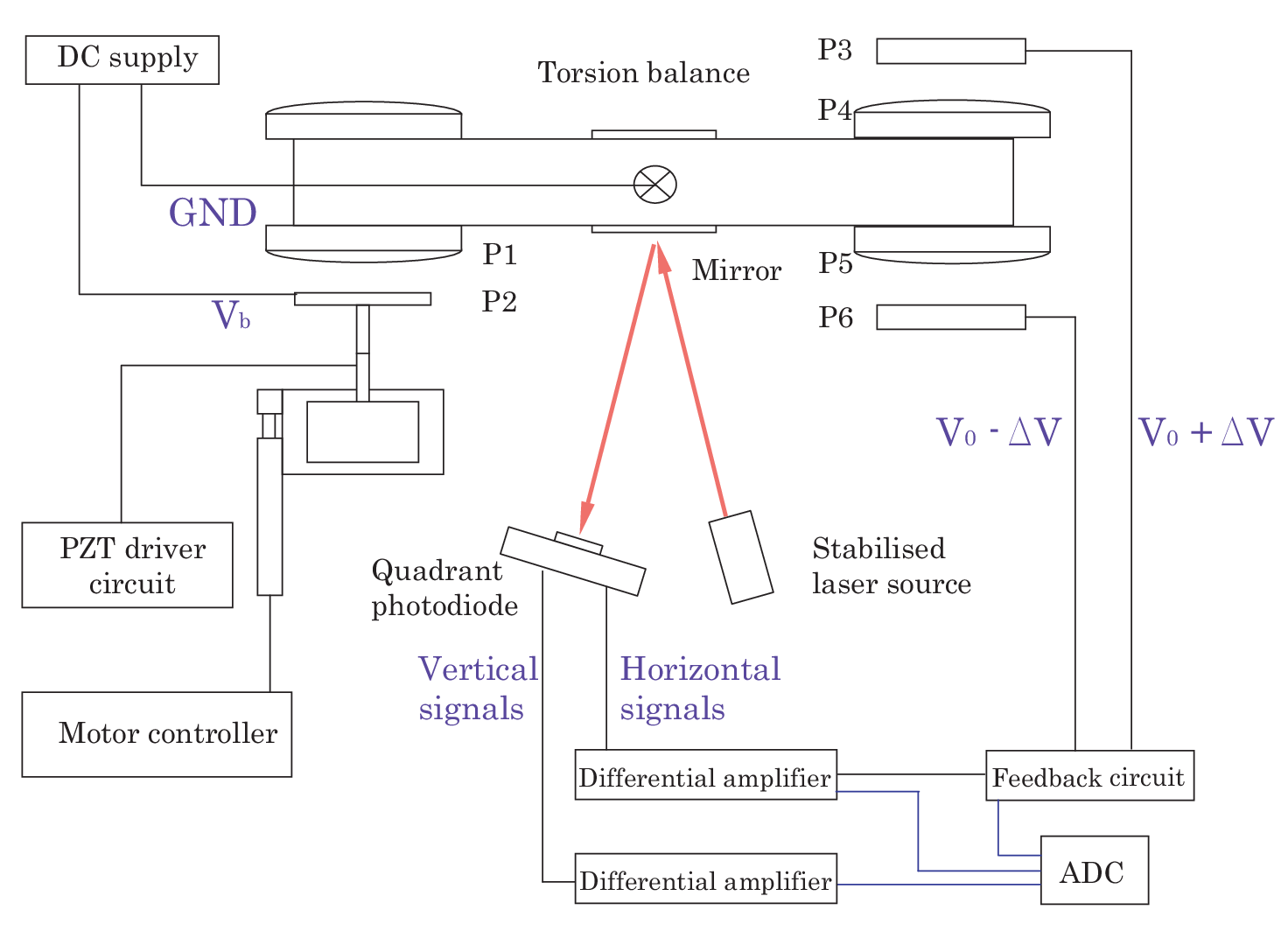} 
\caption{Schematic diagram of the instrument. The red arrows represent the laser light. 
P1 and P2 indicate the spherical lens and the flat plate that produce a macroscopic force, respectively.
P3, P4, P5 and P6 indicate plates forming actuators 
that are used to control the torsion angle. 
Plates P1, P4 and P5 are electrically grounded by the wire of the torsion balance.
A bias voltage $V_b$ can be applied to plate P2. 
Voltages $V_0+\Delta V$ and $V_0-\Delta V$ are applied to plates P3 and P6, respectively, 
during feedback, where $\Delta V$ is the feedback voltage and $V_0=6.0$ V is the bias voltage.
}
    \label{blockdiagram}
    \end{center}
\end{figure}

The instrument is designed to measure the dependence of the force on the distance 
by controlling the distance. 
An optical lever is used to measure the torsion angle, 
and a feedback system is employed to keep the torsion angle constant. 
The instrument is shown in figure \ref{blockdiagram} and consists of a torsion balance, plates for producing a macroscopic force, a feedback system, 
a distance controller and a vacuum chamber.

\subsection{Torsion balance}
\label{torsion balance} 
The torsion balance is a bar, which is made of high-purity (99.999$\%$) copper and is suspended above the centre of gravity 
by a tungsten wire of diameter 60 $\mu$m and length 400 mm.
The wire was annealed in vacuum by passing a current of 400 mA through it 
while the load of the bar was applied, in order to relieve the residual internal stress. 
The torsional period was 165 s and the torsion constant of the wire was $6.1\times10^{-7}$ N$\cdot$ m/rad.
All vibrational motions of the torsion balance were damped using eddy current effects 
produced by placing a strong permanent magnet 2 mm beneath the torsion balance. 
This leads to the quality factor of the torsional mode of 2.5. 
The angular fluctuation of the torsion balance was measured to be 10 $\mu$rad/$\sqrt{\mathrm{Hz}}$ at a frequency of 1 mHz.
This angular fluctuation is equivalent to a force having a root mean square amplitude of 4 pN, which is sufficiently small to measure the Casimir force.

\subsection{Plates for producing a macroscopic force}
\label{plates} 
A spherical lens P1 and a flat plate P2 are used to produce a macroscopic force. 
We utilize optical lenses made of BK7 glass as substrates for both plate P1 and plate P2.  
Plate P1 has a diameter of 40 mm, a thickness of 5 mm, a radius of curvature of 207 mm and the root mean square amplitude of its surface roughness was measured to be 22 nm.
Plate P2 has a diameter of 30 mm, a thickness of 2 mm and the root mean square amplitude of its surface roughness is nominally less than 10 nm.
These surfaces were metalized by evaporation coating a chromium layer with a thickness of 20 nm and then evaporation coating a gold layer with a thickness of 1 $\mu$m. 
Plate P1 was electrically grounded using the bar and the wire.
A bias voltage can be applied between plates P1 and P2.

\subsection{Feedback system} 
\label{feedback system} 
A feedback system is employed to measure macroscopic forces using a null method.
The feedback system is shown in figure \ref{blockdiagram} and consists of a torsion balance, 
an optical lever, analog feedback circuits and two plates used as actuators (P3 and P6).
The torsion angle of the torsion balance is detected using an optical lever, 
and the signal is then fed back to the torsion balance by applying electric forces 
in order to compensate for the deflection of the torsion angle. 

In the optical lever, laser light is incident onto a flat mirror which is attached to the centre 
of the torsion balance, and the reflected light is detected using a quadrant photodiode.
The torsion angle can be obtained by reading out the difference 
between the horizontal signals of the quadrant photodiode. 
However, when the torsion balance undergoes pendulum-like motion due to disturbances such as 
seismic motions and ground tilts, the distance between plates P1 and P2 varies and 
the angle between the wire and the ground changes. 
The tilt angle of the wire can also be obtained by reading out the difference 
between the vertical signals of the quadrant photodiode. 
The output coefficient of the quadrant photodiode in both the horizontal and vertical directions 
is 2.3 mV/$\mu$rad. 

A helium-neon laser was used as the light source for the optical lever. 
The beam from the laser was split into two beams using a polarizing beam splitter, and the intensities of the p and s polarization components were measured using photodiodes. 
The intensity of the laser light was stabilized by controlling the temperature of the laser cavity 
so that the intensities of the two polarization components were equal. 
This stabilization results in an angular sensitivity of $1\times10^{-6}$ rad/$\sqrt{\mathrm{Hz}}$ at 1 mHz.

In the feedback system, the signal of the torsion angle is integrated, filtered and then converted to 
the feedback voltage $\Delta V$ in the analog feedback circuit.
When voltages $V_0+\Delta V$ and $V_0-\Delta V$ are applied to plates 
P3 and P6, respectively, where $V_0$ is the bias voltage, 
an electric force proportional to $\Delta V$ acts on the torsion balance. 
The feedback response was adjusted to give a unity gain frequency of approximately 0.04 Hz 
and a quality factor of the closed loop of 2.5. 
Measuring the feedback voltage enables us to detect forces acting on plates P1 and P2. 
A feedback voltage of 1 mV is equivalent to a force of $1.1\times10^{-12}$ N acting between them.

\subsection{Distance controller}
\label{distance controller} 
The position of plate P2 is controlled coarsely using a motorized stage and finely using a  
piezo translator (PZT). 
The relationship between the displacement and the voltage applied to the PZT was calibrated.
The displacement of the PZT is 11.0 ${\mu}$m when a bias voltage of 100 V was applied. 
The PZT is used only along calibrated routes since their responses exhibit hysteresis and non-linearity.
The direction of plate P2 can be controlled using a manual $\theta$-$\phi$ stage
having a precision of 0.2 mrad before measurements to ensure the spherical lens and the plate are aligned. 

\subsection{Vacuum chamber}
The torsion balance, plates P1--P6 and the motorized stage are set in a vacuum chamber.
The chamber is evacuated using an oil-free scroll pump 
and a turbo-molecular pump prior to taking measurements.
A pressure of the order of 0.1 Pa was maintained during measurements.

\section{Measurements}
\label{measurements} 
In order to estimate the sensitivity to a macroscopic force in the sub-micrometre range, 
measurements were performed in two different locations, Tokyo and Esashi.
The laboratory in Tokyo is located in a basement of a building while 
the laboratory in Esashi is located in a tunnel in a mountainside.
A typical spectrum of the seismic motion in Esashi was measured to be 
$2\times10^{-9}$ m/$\sqrt{\mathrm{Hz}}$ at 1 Hz, which is much smaller than 
that ($1\times10^{-7}$ m/$\sqrt{\mathrm{Hz}}$ at 1 Hz) in Tokyo.
The transfer functions from seismic motions or ground tilts 
to the distance between plates P1 and P2 
were measured in Tokyo by taking advantage of the large seismic motions and ground tilts that occur there.
On the other hand, calibrations of the instrument and measurement of the electric forces 
were performed in Esashi.

\subsection{Noise estimation} 
We measure the force dependence on the distance between plates P1 and P2 
by varying the distance in constant step intervals. 
For each step, we use an averaging time of 400 s and a dead time of 200 s. 
The fluctuations in the distance between the plates during the averaging time 
give rise to systematic errors in the force since the Casimir force is a non-linear function of distance.
Here we estimate the distance fluctuations due to 
seismic motions, ground tilts, angular fluctuations and thermal fluctuations. 

First of all, the distance fluctuations due to seismic motions are estimated as follows.
It is assumed that the centre of mass of the torsion balance 
has translational motion with an amplitude of $\tilde{x_p}(\omega)$ 
caused by translational seismic motion $\tilde{x_s}(\omega)$. 
By defining and solving the equation of motion for $\tilde{x_p}(\omega)$,
the distance between plates P1 and P2 $\tilde{d}(\omega)$ is given by

\begin{equation}
\tilde{d}(\omega)=\tilde{x_p}(\omega)-\tilde{x_s}(\omega)
=\frac{(\frac{\omega}{\omega_0})^2}{1-(\frac{\omega}{\omega_0})^2+i\frac{1}{Q}(\frac{\omega}{\omega_0})}\tilde{x}_s(\omega),
\label{distance}
\end{equation}
where $Q$ is the quality factor of the translational motion and $\omega_0$ is the resonant frequency of the translational motion.
Let $H(\omega)$ be the transfer function defined 
as $H(\omega)\equiv\tilde{\theta}(\omega)/\tilde{x_s}(\omega)$.
The tilt of the instrument $\tilde{\theta}(\omega)$ is simply given by
$\tilde{\theta}(\omega)=\tilde{d}(\omega)/l$, where $l$ is the length of the wire.
Therefore the transfer function  $H(\omega)$ is given by 
\begin{equation}
\label{tftf}
|H(\omega)|=\biggl|\frac{1}{l}\frac{(\frac{\omega}{\omega_0})^2}{1-(\frac{\omega}{\omega_0})^2+i\frac{1}{Q}(\frac{\omega}{\omega_0})}\biggr|
=\frac{1}{l}\frac{(\frac{\omega}{\omega_0})^2}
{\sqrt{\{1-(\frac{\omega}{\omega_0})^2\}^2+(\frac{1}{Q}\frac{\omega}{\omega_0})^2}}.
\end{equation}

\begin{figure}[h]
    \begin{center}  
\includegraphics[width=12cm]
{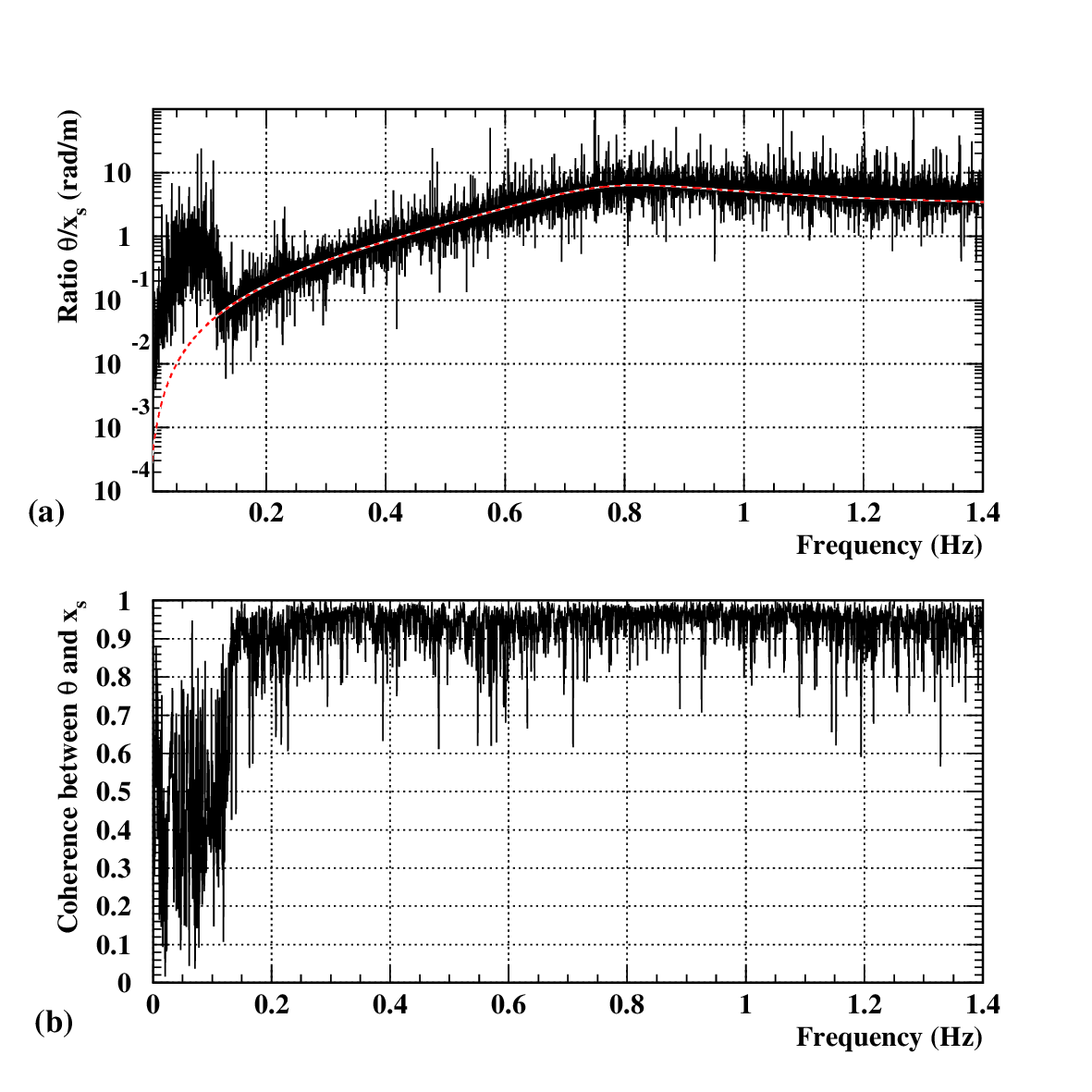}
    \caption{(a) Ratio and (b) coherence 
between the translational seismic motion $\tilde{x_p}(\omega)$ and the tilt angle of the wire $\tilde{\theta}(\omega)$.
The ratio was fitted using equation (\ref{tftf}) with a single value for $Q$, and $Q$ was thus determined to be 2.0.
The red dashed line is the fitted function. }
    \label{data1019tf}
    \end{center}
\end{figure}

\begin{figure}[h]
    \begin{center}  \includegraphics
[width=12cm]{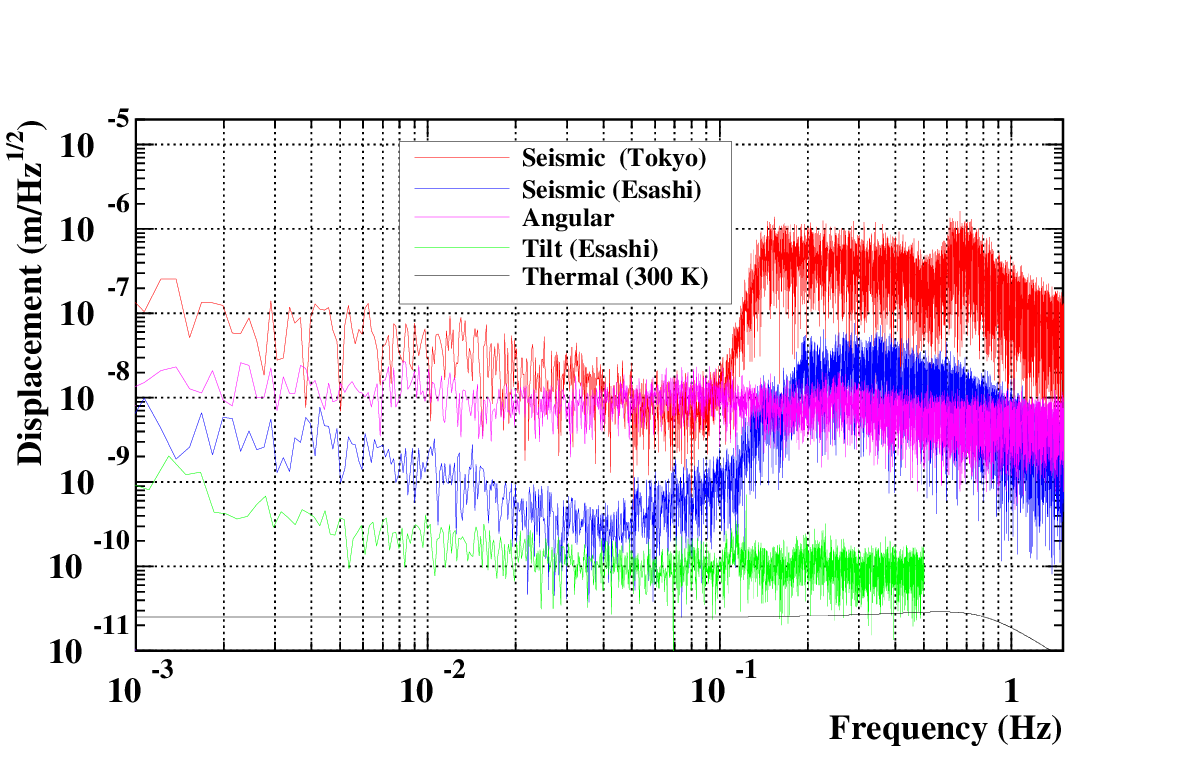}
    \caption{
Expected spectrum of the distance fluctuations between plates P1 and P2. 
The red and blue lines represent the seismic fluctuations in Tokyo and Esashi, respectively. 
The purple, green and black lines represent the suppressed angular fluctuations, the tilt fluctuations in Esashi 
and the thermal fluctuations at 300 K, respectively.}
    \label{data1106sep}
    \end{center}
\end{figure}

In order to estimate the validity of the above model, 
we measured the translational seismic motion 
and the tilt angle of the wire simultaneously in Tokyo. 
The ratio of the translational seismic motion to the tilt angle of the wire 
is compared with the transfer function $H(\omega)$ in figure \ref{data1019tf}.
This figure shows that the ratio is consistent with the transfer function 
in the frequency range between 0.1 and 1.4 Hz, in which the coherence is as high as $\sim1$.
In the frequency range below 0.1 Hz, 
we confirmed that the coherence is low since the detection noise of the optical lever is 
larger than the signal of the tilt. 
Thus the transfer model in equation (\ref{tftf}) can be used to 
estimate the distance fluctuations due to seismic motions. 
We then estimate the spectra of the distance fluctuations  
due to seismic motions in Tokyo and Esashi, 
assuming that $Q$ is set to unity by controlling the position of the damping magnet. 
The results are shown in figure \ref{data1106sep}. 

Ground tilts also cause the distance between plates P1 and P2 to fluctuate.
We measured the ground tilts and estimated the distance fluctuations due to this effect 
in Esashi and Tokyo.
The laboratory in Tokyo was located in the basement of a building and in this location
the daily variation in the ground tilt had an amplitude of a few $\mu $rads.  
The laboratory in Esashi was located in a tunnel and in this location the ground tilt had an amplitude of several tens of nrads 
with a period of half a day and we confirmed that this ground tilt was caused by the earth tide.
The amplitude of the ground tilt in Esashi was two orders of magnitude smaller than that in Tokyo.
The estimated spectrum of the distance fluctuation between plates P1 and P2 
due to the ground tilt is shown in figure \ref{data1106sep}.
The distance fluctuation due to the ground tilt in Esashi is smaller than 
that due to seismic motions, and is negligible for measurements of the Casimir force. 

The thermal fluctuations of the tilt angle of the wire must be taken into account, 
since measurements were done at room temperature ($\sim$300 K).
The thermal fluctuation of the distance between plates P1 and P2 
$\tilde{d}_{T}(\omega)$ is given by 
\begin{equation}
|\tilde{d}_{T}(\omega)|=\sqrt{\frac{4k_BT}{mQ}\frac{\omega_0}{(\omega^2-\omega_0^2)^2+\omega_0^2\omega^2/Q^2}},
\label{dthermal}
\end{equation}
where $m$ is the mass of the torsion balance, $k_B$ is the Boltzmann constant and $T$ is the temperature. 
As shown in figure \ref{data1106sep}, 
the thermal fluctuations of the distance at 300 K are one to three orders of magnitude smaller 
than those due to seismic motions in Esashi, and thus it cannot be a dominant noise source. 

While a feedback system is employed 
to keep the torsion angle fixed, 
angular fluctuations dominantly caused by the noise of the optical lever still remain. 
Figure \ref{data1106sep} shows that
the fluctuations in the distance due to the residual angular fluctuation 
are dominant in the frequency range from 1 mHz to 0.1 Hz in Esashi.

By integrating these estimated spectra over the range from 1 mHz to 1 Hz, 
the root mean square amplitude of the distance fluctuations 
due to all effects in Esashi was determined to be $d_{rms}$ = 18 nm.

\subsection{Sensitivity} 
First, we estimate the sensitivity to macroscopic forces 
by measuring the electric force which was intentionally applied between plates P1 and P2. 
This electric force was used as a reference since the dependence of the electric force on distance can be calculated with high accuracy. 
If a bias voltage $V_{b}$ is applied between a flat plate and a spherical lens having a radius of curvature of $R$, 
the electrostatic force $F_e(V_{b},d)$ is expressed as 
\begin{equation} 
F_{e}(V_b,d)=\frac{{\pi}{\epsilon}_{0}R(V_{c}-V_{b})^{2}}{d} \quad  \mathrm{for} \quad d\ll R ,
\label{electric2}
\end{equation}
where $V_{c}$ is the contact potential difference and ${\epsilon _0}$ is the dielectric constant. 
The electric contact potential difference between two metallic surfaces is 
caused by the different work functions of different metals or the distribution of surface charges.
If the bias voltage is varied with an amplitude of $V_s$ at a certain distance ($d=d_0$), 
the variation in the electrostatic force $\Delta F_{v}(V_b)$ is given by 
\begin{eqnarray*}
\Delta F_{v}(V_b)=F_{e}(V_{b}+V_s,d_0)-F_{e}(V_{b},d_0)
\end{eqnarray*}
\begin{eqnarray}
=\frac{2V_s{\pi}{\epsilon}_{0}R}{d_0}V_{b}+\frac{{\pi}{\epsilon}_{0}R}{d_0}(V_s^{2}-2V_{s}V_{c}).
\label{elefit}
\end{eqnarray}
In a similar way, if the distance is varied with a constant step $s$ 
at a constant bias voltage ($V_b=V_{0}$), 
the variation in the electric force $\Delta F_{s}(d)$ is given by
\begin{equation} 
\Delta F_{s}(d)=F_{e}(V_{0},d)-F_{e}(V_{0},d+s)=\frac{{\pi}{\epsilon}_{0}R(V_{c}-V_{0})^{2}s}{d(d+s)}.
\label{dfele}
\end{equation}

\begin{figure}[h!]
    \begin{center}  
\includegraphics[width=12cm]{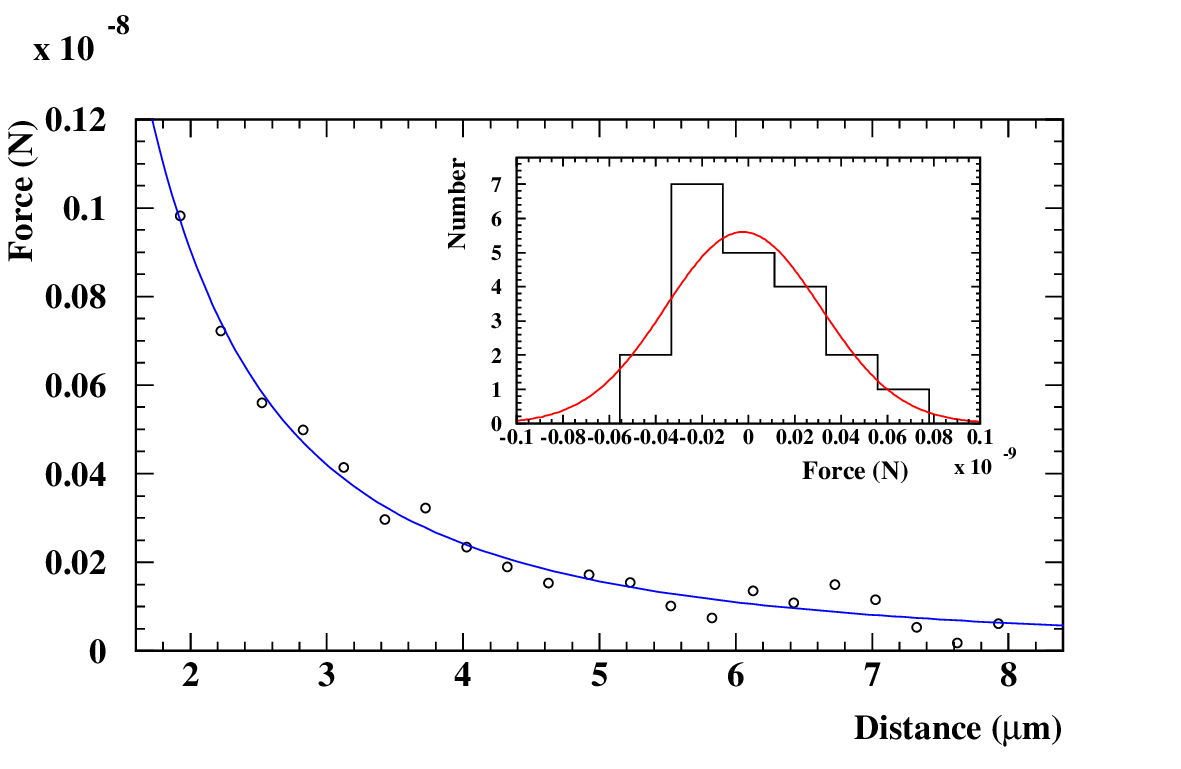}
    \caption{$\Delta F_{s}(d)$ as a function of distance. 
The blue line is the function fitted to equation (\ref{dfele}). 
Inset: distribution of the deviation between the data and equation (\ref{dfele}).
The standard deviation of the distribution is determined to be 
$\sigma_F=3.4\times10^{-11}$ N.}
@@\label{dfdx0311c1b}
    \end{center}
\end{figure}

In order to determine the absolute distance between plates P1 and P2, 
we measured $\Delta F_{v}(V_b)$ when the bias voltage 
was changed in steps of 10 mV at a certain distance. 
By fitting the data to equation (\ref{elefit}) with fitting parameters $V_c$ and $d_0$, 
we obtained $V_c=82.6\pm0.9$ mV and $d_0=1.601\pm0.013$ ${\mu}$m.
We then measured $\Delta F_{s}(d)$ 
by bringing plate P2 close to plate P1 
with a constant step of 0.3 $\mu$m from 8.1 $\mu$m, as shown in figure \ref{dfdx0311c1b}. 
The absolute distances between plates P1 and plate P2 were determined 
from $d_0$ and the voltages applied to the PZT. 
The standard deviation between the data and equation (\ref{dfele}) was 
determined to be $\sigma_F=3.4\times10^{-11}$ N.

We treated the error in the absolute distance and the distance fluctuations during measurements 
as systematic errors.
In the case when the variation in the Casimir force $\Delta F_{cas}(d)=F_{cas}(d)-F_{cas}(d+s)$
is measured at $d=1.0$ $\mu$m with $s=0.3$ $\mu$m, 
the systematic error of the force $\sigma_{d0}$ 
propagated from the error of the absolute distance $\Delta d_0=13$ nm is given by
\begin{equation}
\sigma_{d0}=\frac{\partial \Delta F_{cas}(d)}{\partial d} \Delta d_0
=1.4\times10^{-11}  \; \mathrm{N} \quad \mathrm{for} \quad d=1.0 \;  \mu\mathrm{m}.
\label{sigmad0}
\end{equation}

Furthermore we estimated the systematic error of the force 
due to the distance fluctuations during measurements, by using the Monte Carlo method.
Assuming the distance fluctuations follow a Gaussian distribution with a standard deviation of $d_{rms}$ = 18 nm, 
the systematic error of the force is estimated to be  
$\sigma _{drms}=8.3\times 10^{-13}$ N at $d=1.0$ $\mu$m. 
In the case when $\Delta F_{cas}(d)$ is measured 100 times at the distance, 
the statistical error of the force is estimated to be $\sigma_{stat}=3.4\times10^{-12}$ N, 
which is smaller than the systematic error due to the error of the absolute distance.

Since these errors might be correlated, 
we conservatively estimate the total error of the force to be
$ {\Delta}F=\sigma_{stat}+\sigma _{d0}+\sigma _{drms}=1.8\times 10^{-11}$ N.
This means that variations in the Casimir force $\Delta F_{cas}(d)$ 
can be measured with a relative error of 6$\%$ at $d$ = 1.0 $\mu$m, including systematic errors.

Now, the sensitivity to a non-standard force is estimated 
by assuming that the potential of the non-standard force is parameterized by the Yukawa potential as follows:
\begin{equation} 
\label{yukawa}
V_{Y}=-\alpha G \int \mathrm{d}\mathbf{r}_1 \int \mathrm{d}
\mathbf{r}_2 \frac{\rho_1(\mathbf{r}_{1}) \rho_2(\mathbf{r}_{2})}{\mathbf{r}_{12}}\mathrm{e}^{-\mathbf{r}_{12}/\lambda },
\end{equation}
where $G$ is Newton's constant, $\rho_1$ and $\rho_2$ are the densities of the two materials, 
$\mathbf{r}_{12}$ is the distance between the materials, 
$\alpha$ is the strength of a non-standard force with respect to Newtonian gravity 
and $\lambda$ is the range parameter. 
In this case, the non-standard force 
between a spherical lens and a flat plate when $\lambda\leq d$ 
is approximately given by 
\begin{equation}
F_{Y}(d)= 
{\alpha }G4{\pi }^2{\lambda }^3\mathrm{exp}(-d/\lambda)Rg(\lambda)^2,
\label{fyukawa}
\end{equation}
where $g(\lambda)={\rho_1}\mathrm{exp}(-(\Delta_1+\Delta_2)/\lambda)+{\rho_2}\mathrm{exp}(-\Delta_2/\lambda)+{\rho_3}$
is a geometrical form factor, ${\rho_i}$ ($i=1,2,3$) is the density of the $i$th layer and $\Delta_j$ ($j=1,2$) is the thickness of the $j$th layer \cite{Bordag2}.
In our configuration, ${\rho_1}$ is the density of BK7, ${\rho_2}$ is the density of chromium, ${\rho_3}$ is the density of gold, $\Delta_1$ is
the thickness of chromium and $\Delta_2$
is the thickness of gold.

If the variation in the non-standard force $F_{Y}(d)$, 
when the distance is changed with a constant step $s$, 
is measured and if the non-standard force is lower limited by the force error ${\Delta} F$, 
then the parameter is limited as 
\begin{equation}
\alpha \leq \frac{{\Delta}F}{G4{\pi }^2{\lambda }^3R
g(\lambda)^2}\frac{\mathrm{exp}(d/\lambda)}{1-\mathrm{exp}(-s/\lambda)}.
\label{hutou}
\end{equation}

\begin{figure}[h]
 \begin{center}  
\includegraphics[width=12cm]
{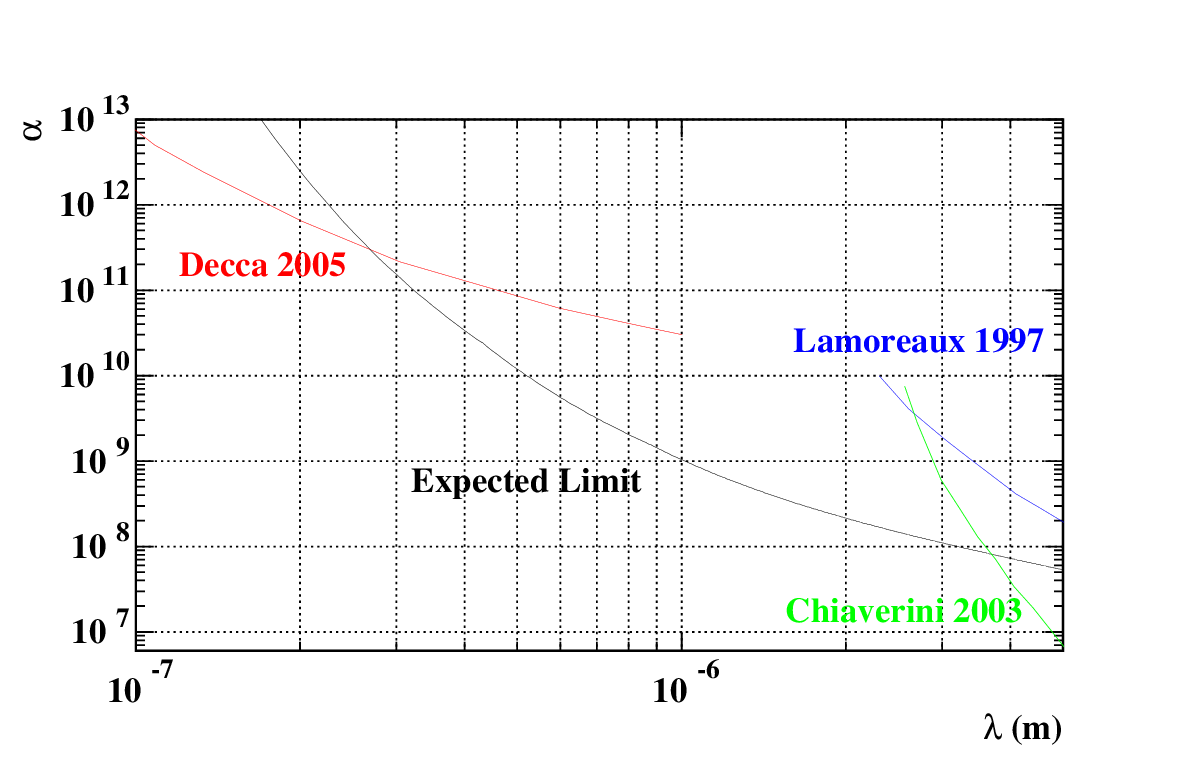}
    \caption{Limits on $\alpha$ as a function of $\lambda$ assuming that the non-standard force has a Yukawa-type potential. 
The black curve shows the expected limit in this experiment. 
The region above the curves is excluded. 
It is assumed that the data are measured 100 times at each distance $d=0.8, 1.1, 1.4$ and $1.7$ $\mu $m 
with $\Delta d$ = 0.3 $\mu$m.
The red, green and blue curves show the limits obtained 
for a micro-electro-mechanical system \cite{Decca2},  
a microcantilever \cite{chiaverini} and a torsion balance \cite{Lamoreaux1}, respectively.}
    \label{alphasence06}
    \end{center}
\end{figure}

When the data are measured 100 times at each distance $d$=0.8, 1.1, 1.4 and 1.7 $\mu$m 
with $s=0.3$ $\mu$m, the total error of the force is estimated to be   
${\Delta}F=4.5\times10^{-11},1.3\times10^{-11}, 6.6\times10^{-12}$ and $4.7\times10^{-12}$ N, 
respectively.
The upper limit to the non-standard force is estimated from equation (\ref{hutou}) 
and is shown in figure \ref{alphasence06}.

\section{Conclusions}
\label{sec:conclusions} 
We have developed an instrument that employs a torsion balance for 
probing a non-standard force in the sub-micrometre range. 
High sensitivity is achieved by using a torsion balance 
that has a long torsional period, strong magnetic damping of all vibrational motions 
and a feedback system that employs an optical lever. 
We have estimated the distance fluctuations due to 
seismic motions, tilt motions, residual angular fluctuations and thermal fluctuations. 
The estimations indicate that the seismic motion is the most dominant noise source of  
the distance fluctuations even for the stable site at Esashi. 
The root mean square amplitude of the total distance fluctuation was estimated to be 18 nm.
An error in the absolute distance of 13 nm was obtained by measuring 
the dependence of the electric force on the bias voltage. 
We also estimated the statistical error to be $3.4\times 10^{-12}$ N 
by measuring the dependence of the electric force on distance. 
From this systematic study, 
we conclude that the Casimir force can be measured with a relative error of 6$\%$ at $d=1.0$ $\mu$m 
and the instrument could be used to obtain a stringent limit on the non-standard force in the 0.3--3.7 $\mu$m range. 

\ack 
We wish to thank Dr. K. Yamamoto and Prof. T. Tsubokawa for their valuable discussions on seismic motion and ground tilt.
This study was supported in part by the Earthquake Research Institute Cooperative Research Program 
and cooperative use of Mizusawa VERA Observatory of the National Astronomical Observatory of Japan.

\end{document}